\documentclass[reprint,twocolumn,footinbib,longbibliography,superscriptaddress]{revtex4-1}

\usepackage{graphicx,dcolumn,bm,amsmath,amssymb,amsfonts,lmodern,times,amsthm,mathtools,physics,mathrsfs,hyperref,physics,hyperref,url}

\newcommand{\jp}{{j^\prime}}
\newcommand{\iprime}{{i^\prime}}
\newcommand{\Metabo}{{\mathcal{M}}}
\newcommand{\Const}{{\mathcal{C}}}
\newcommand{\Reac}{{\mathcal{R}}}
\newcommand{\Path}{{\mathcal{P}}}
\newcommand{\Obj}{{\mathcal{O}}}
\newcommand{\Excess}{{\mathcal{E}}}

\newcommand{\Income}{{{\bf I}}}
\newcommand{\flux}{{{\bf f}}}
\newcommand{\f}{{{\bf f}}}
\newcommand{\vflux}{{{\bf v}}}
\usepackage{color}

\begin{document}
\title{Linear Response Theory of Evolved Metabolic Systems}
\author{Jumpei F. Yamagishi}
\author{Tetsuhiro S. Hatakeyama}
\affiliation{Department of Basic Science, The University of Tokyo, 3-8-1 Komaba, Meguro-ku, Tokyo 153-8902, Japan}

\begin{abstract}
Predicting cellular metabolic states is a central problem in biophysics. Conventional approaches, however, sensitively depend on the microscopic details of individual metabolic systems. In this Letter, we derived a universal linear relationship between the metabolic responses against nutrient {conditions} and metabolic inhibition, with the aid of a microeconomic theory. The relationship holds in arbitrary metabolic systems as long as the law of mass conservation stands, as supported by extensive numerical calculations. It offers quantitative predictions without prior knowledge of systems.
\end{abstract}

\maketitle

Metabolism is the physicochemical basis of life. Understanding its behavior has been a major goal of biophysics~\cite{ilker2019modeling,kar2003collapse,yamagishi2020advantage,zwicker2017growth}. 
At the same time, {the} prediction of cellular metabolic states is a central problem in biology. In particular, prediction of the responses of metabolic systems against environmental variations or experimental operations is essential for manipulating metabolic systems to the desired states in both life sciences and in applications such as matter production in metabolic engineering~\cite{stephanopoulos1998metabolic} and the development of drugs targeting cellular metabolism~\cite{murima2014targeting,drug-induced-hyperthermia,martinez2017cancer}. 

Previous studies have mainly {attempted} to predict the metabolic responses by predicting the metabolic states before and after perturbations, and {they require} building an \textit{ad hoc} model for each specific metabolic system. 
In systems biology, constraint-based modeling (CBM) has often been used to predict the cellular metabolic states~\cite{palsson2015systems,klipp2016systems,warren2007duality}. 
In this method, the intracellular metabolic state is predicted by solving an optimization problem of models of metabolic systems, including a detailed description of each metabolic reaction. 
To construct the optimization problem, metabolic systems of cells are assumed to be optimized through (sometimes artificial) evolution for some objectives~\cite{heinrich1998modelling,klipp2016systems,palsson2015systems}, e.g., maximization of the growth rate in reproducing cells such as cancer cells and microbes~\cite{OMbook} and maximization of the production of some molecules in metabolically engineered cells~\cite{portnoy2011adaptive}. 
Indeed, metabolic systems of reproducing cells exhibit certain ubiquitous phenomena across various species, and those phenomena can be explained as a result of optimization under physicochemical constraints~\cite{OMbook}. 
Although the assumption of optimal metabolic regulation seems acceptable, knowing the true objective function of cells, which is essential to making a model for CBM, remains nearly impossible. 
Besides, even with remarkable progress in omics research, fully reconstructing metabolic network models for each individual species or cell of interest is still a challenge. 
{Moreover}, the numerical predictions are sensitive to the details of the concerned constraints and the objective functions selected~\cite{bonarius1997flux,raman2009flux,schnitzer2022choice,Heinemann-Gibbs2019}. 
Therefore, new methods independent of the details of metabolic systems are required.

Instead of metabolic states themselves, here, we focus on the responses of metabolic systems to perturbations. At first glance, such prediction is seemingly more difficult than predicting the cellular metabolic states because it seems to require information not only on the steady states but also on their neighborhoods. 
However, from another perspective, to predict only the metabolic responses, we may need to understand the structure of only a limited part of the state space of feasible metabolic states. In contrast, we must seek the whole space to predict the metabolic states themselves. 
If optimization through evolution and some physicochemical features unique to metabolic systems constrain the behavior in the state space, there might be universal features in the responses of metabolic systems to perturbations, independent of system details, as in the linear response theory in statistical mechanics~\cite{onsager1931reciprocal,kubo1957statistical,green1954markoff}. 

In this Letter, we demonstrate a universal property of intracellular metabolic responses in the optimized metabolic regulation, using a microeconomic theory~\cite{Varian,Lancaster1966,yamagishi2021microeconomics}. By introducing a microeconomics-inspired formulation of metabolic systems, we can take advantage of tools and ideas from microeconomics such as the Slutsky equation that describes how consumer demands change in response to income and price. We thereby derive quantitative relations between the metabolic responses against nutrient abundance and those against metabolic inhibitions, such as {the} addition of metabolic inhibitors and leakage of intermediate metabolites; the former is easy to measure in experiments while the latter may not be. 
The relations universally hold independent of the details of metabolic systems as long as the law of mass conservation holds. Our theory is applicable to any metabolic {system} and will provide quantitative predictions on the intracellular metabolic responses without detailed prior knowledge of microscopic molecular mechanisms and cellular objective functions.

\begin{figure}[tb]
    \centering\includegraphics[width=\linewidth]{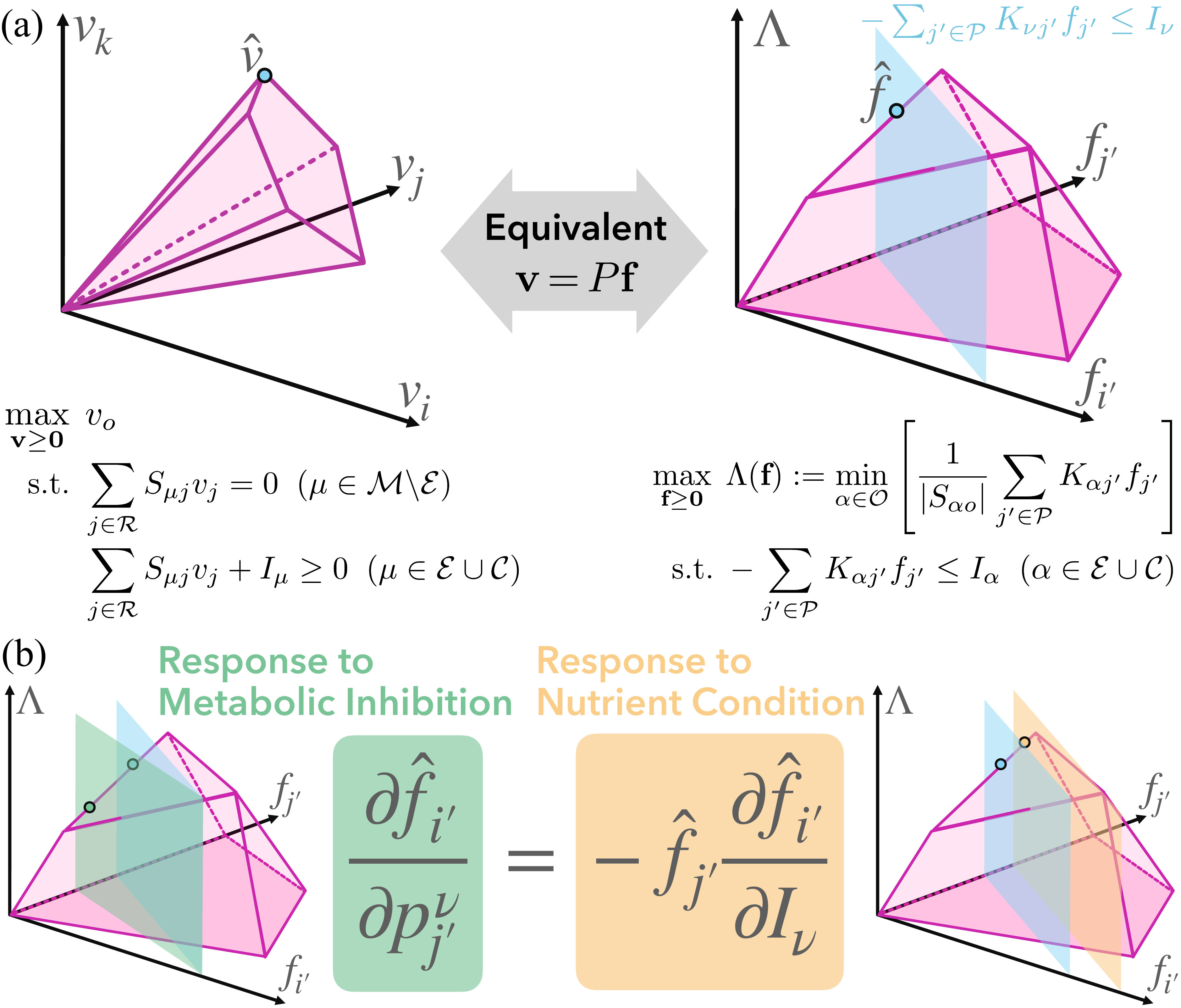}
    \caption{
    Schematic illustration. 
    (a) (left) Metabolic CBM formulation with reaction fluxes $\vflux$ as variables. The solution subspace (convex set of possible allocations), called {the} flux cone, is shown in pink. 
    (right) Microeconomic formulation with pathway fluxes $\flux$ as variables and an objective flux $\Lambda$. The pink area in $\flux$-plane (bottom surface) represents the solution subspace, whereas the blue plane vertical to $\flux$-plane is the budget constraint for a component $\nu$. 
    The blue points $\hat{\vflux}$ and $\hat{\flux}$ represent the optimized fluxes of reactions and pathways, respectively. 
    Given $\vflux\,{=}\,P\flux$ with pathway matrix $P$, both formulations are equivalent optimization problems (see Supplemental Material (SM), Sec.~S1 for details {and Sec.~S2 and Fig.~\ref{fig:SI_toy_model} for a simple example}). 
    (b) Liner relation between the metabolic responses against changes in nutrient conditions (yellow) and those against metabolic inhibitions (green) [Eq.~\eqref{eq:Slutsky_Metabolism}). 
    }\label{fig:fig1}
\end{figure}

\textit{Microeconomic formulation of metabolic regulation.---}
We first provide a microeconomic formulation of optimized metabolic regulation, which is equivalent to linear programming problems in CBM (Fig.~\ref{fig:fig1}). 

We denote the set of all chemical species (metabolites) and that of all constraints by $\Metabo$ and $\mathcal{C}$, respectively. 
$\mathcal{C}$ can reflect every type of {constraints} such as the allocation of proteins~\cite{scott2011bacterial,OM}, intracellular space~\cite{vazquez2010catabolic}, membrane surfaces~\cite{memRealEstate2}, and Gibbs energy dissipation~\cite{Heinemann-Gibbs2019} as well as the bounds of reaction fluxes.  

In the microeconomic formulation, variables to be optimized are the fluxes of metabolic pathways, whereas they are fluxes of reactions in usual CBM approaches; a metabolic pathway is a linked series of reactions and thus comprises multiple reactions. The sets of reactions and pathways are denoted by $\Reac$ and $\Path$, respectively. 
Let us then consider two stoichiometry matrices for reactions and pathways, $S$ and $K$, respectively (see also SM, Table~\ref{table:Symbols}). {For chemical species $\alpha\,(\in\Metabo)$,} $|S_{\alpha i}|$ represents the number of units of species $\alpha$ produced {if} $S_{\alpha i}\,{>}\,0$ and consumed {if} $S_{\alpha i}\,{<}\,0$ in reaction $i$; whereas if $\alpha$ {denotes} a constraint ($\alpha\in\Const$), $S_{\alpha i}$ is usually negative and $|S_{\alpha i}|$ represents the number of units of constraint $\alpha$ required for reaction $i$. The stoichiometry matrix $K$ for metabolic pathways $\Path$ is also defined similarly. Throughout the Letter, we use indices with primes such as $\iprime$ to denote pathways and those without primes such as $i$ to denote reactions, and $|S_{\alpha i}|$ and $|K_{\alpha \iprime}|$ are called input (output) {stoichiometric} coefficients of reaction $i$ and pathway $\iprime$, respectively, if $S_{\alpha i}$ and $K_{\alpha \iprime}$ are negative (positive).

Cells are assumed to maximize the flux of some objective reaction $o\,(\in\Reac)$ such as biomass synthesis in reproducing cells and ethanol or ATP synthesis in metabolically engineered cells. 

{We define the set of the species consumed in and the components required for reaction $o$ as objective components $\Obj\,(\subset\Metabo\cup\Const)$, and thus $S_{\alpha o}$ for each objective component $\alpha\,(\in\Obj)$ is negative.} 
Because the reactants of a reaction cannot be compensated for each other due to the law of mass conservation{~\cite{yamagishi2021microeconomics,liao2020modeling,roy2021unifying}, the flux of objective reaction $o$, i.e., the objective function, is limited by the minimum available amount of objective components $\Obj$ as follows:}
\begin{eqnarray} \label{eq:growth}
\Lambda(\flux) := \min_{\alpha\in\Obj}\left[\frac{1}{-S_{\alpha o}}\left( \sum_{j^\prime\in\Path} K_{\alpha j^\prime}f_{j^\prime} +I_\alpha \right) \right],
\end{eqnarray}
where $\flux \,{=}\,\{ f_\iprime \}_{\iprime\in\Path}$ represents the fluxes of metabolic pathways. 
The arguments of the {above} min function represent biologically different quantities: if $\alpha$ is a species ($\alpha\in\Metabo$), $I_\alpha$ is {its} intake flux and $\sum_{j^\prime} K_{\alpha j^\prime}f_{j^\prime}$ represents {its} total production rate, while if $\alpha$ is a constraint ($\alpha\in\Const$), $I_\alpha$ is the total capacity for constraint $\alpha$ and $\sum_{j^\prime} K_{\alpha j^\prime}f_{j^\prime} +I_\alpha$ is the amount of $\alpha$ that can be allocated to the objective reaction.

{The optimized solution $\hat{\flux}$ is determined as a function of $K$ and $\textbf{I}$ with the following constraints for the available pathway fluxes $\flux$:} 
\begin{eqnarray}\label{eq:n_goods-m_objectives}
-\sum_{j^\prime\in\Path}K_{\alpha j^\prime}f_{j^\prime}\leq I_\alpha. \quad{(\alpha\in\mathcal{E}\cup\Const)}
\end{eqnarray}
{Here, $\mathcal{E}\,(\subset\Metabo)$ denotes the set of exchangeable species that are transported through the cellular membrane. That is, the above constraints reflect that the total consumption of species cannot exceed their intakes.} 
If species $\alpha$ is produced by objective reaction $o$, the intake effectively increases and $S_{\alpha o}\Lambda$ is added to the right-hand side of Eq.~\eqref{eq:n_goods-m_objectives}, although this is not the case for most species. 

This optimization problem~(\ref{eq:growth}-\ref{eq:n_goods-m_objectives}) can be interpreted as a microeconomic problem in the theory of consumer choice~\cite{Varian,Lancaster1966,yamagishi2021microeconomics}, considering $\Lambda(\flux)$ as the utility function. By focusing on an arbitrary component $\nu$, one of inequalities~\eqref{eq:n_goods-m_objectives} serves as the budget constraint for $\nu$ if $K_{\nu j^\prime}\,{\leq}\,0$ for all pathways $j^\prime$, while the remaining inequalities in Eq.~\eqref{eq:n_goods-m_objectives} then determine the solution space [Fig.~\ref{fig:fig1}(a)]: for example, if we choose glucose as $\nu$, the corresponding inequality in Eq.~\eqref{eq:n_goods-m_objectives} represents carbon allocation. 
Here, the maximal intake $I_\nu$ of $\nu$ corresponds to the income, and the input {stoichiometric} coefficient for each pathway, $p_{j^\prime}^\nu:\,{=}\,-K_{\nu j^\prime}$, serves as the price of pathway $j^\prime$ in terms of $\nu$.

\textit{Relation between responses of pathway fluxes to nutrient abundance and metabolic inhibition.---}
Because Eqs.~(\ref{eq:growth}-\ref{eq:n_goods-m_objectives}) can be interpreted as {a microeconomic} optimization problem, we can apply and generalize the Slutsky equation in {the theory of consumer choice~\cite{Varian}. The equation} shows the relationship between changes in the optimized demands for goods in response to income and price. 
In metabolism, it corresponds to the relationship between the responses of optimal pathway fluxes $\hat{\flux}$ (see {SM, Sec.~S4} for derivation):
\begin{eqnarray}\label{eq:Slutsky_Metabolism}
\frac{\partial \hat{f}_{i^\prime}(K,{\bf I})}{\partial p^\nu_{j^\prime}} = -\hat{f}_{j^\prime}(K,{\bf I})\frac{\partial \hat{f}_{i^\prime}(K,{\bf I})}{\partial I_\nu}.
\end{eqnarray}
The right-hand side represents the responses of pathway $\iprime$ against increases in $I_\nu$, whereas the left-hand side represents those against metabolic inhibitions in pathway $\jp$ because the metabolic price $p^\nu_\jp\,{=}\,-K_{\nu \jp}$ quantifies the inefficiency of conversion from substrate $\nu$ to endproducts in pathway $\jp$~\cite{yamagishi2021microeconomics}.

\begin{figure}[tb]
    \centering\includegraphics[width=\linewidth]{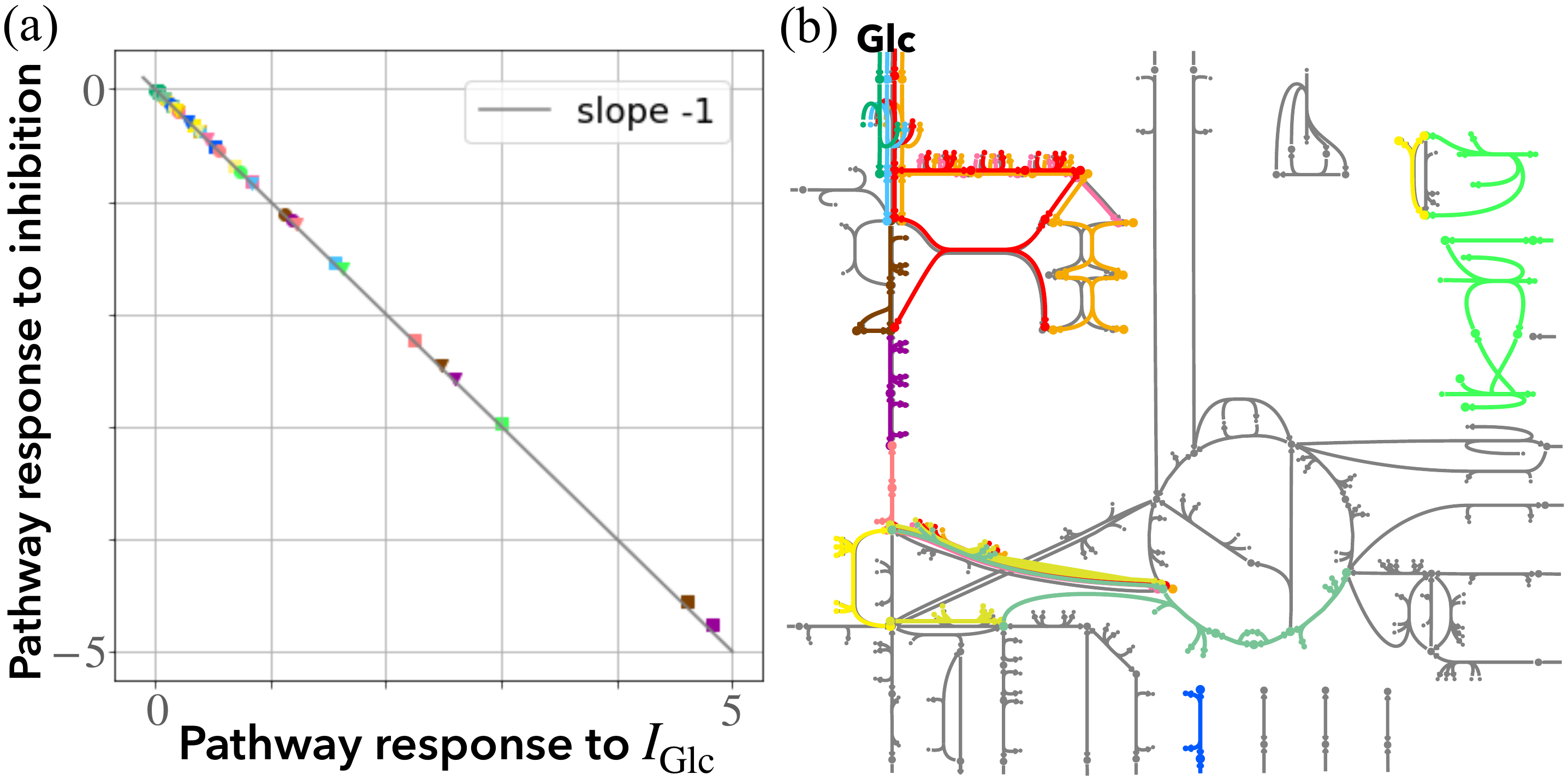}
    \caption{Responses of the optimized pathway fluxes ${\bf \hat{f}}$. 
    (a) Responses to metabolic inhibitions, $\Delta \hat{f}_\iprime(\Delta K_{\mathrm{Glc}, \jp}) / \Delta p^\mathrm{Glc}_\jp$, are plotted against the nutrient responses, $\hat{f}_\jp \Delta \hat{f}_\iprime(\Delta I_\mathrm{Glc}) / \Delta I_\mathrm{Glc}$. 
    All different shapes and colors of markers represent different $\iprime$ and $\jp$, respectively. 
    $I_\mathrm{Glc}\,{=}\,5\,\mathrm{[mmol/gDW/h]}$. 
    (b) $13$ active extreme pathways, computed using efmtool~\cite{efmtool}, are shown. Colors correspond to those of the markers for manipulated pathways $\jp$ in panel (a). The whole metabolic network of the \textit{E. coli} core model is shown in gray. 
    }\label{fig:fig2}
\end{figure}

{The derivation of Eq.~\eqref{eq:Slutsky_Metabolism} relies solely on} the law of mass conservation{, i.e.,} the reactants of a reaction cannot be compensated for each other. 
Because the law of mass conservation stands in every chemical reaction, the relation~\eqref{eq:Slutsky_Metabolism} of the two measurable quantities must hold in arbitrary metabolic systems as long as their metabolic regulation is optimized for a certain objective. 
In particular, the case $\iprime\,{=}\,\jp$ will be useful: it {indicates} that {measuring} the responses of a pathway flux to changes in the nutrient environment provides quantitative predictions of the pathway's responses to metabolic inhibition or activation, and vice versa. 

To confirm the validity of Eq.~\eqref{eq:Slutsky_Metabolism}, we numerically solved the optimization problems~(\ref{eq:growth}-\ref{eq:n_goods-m_objectives}) with pathway fluxes $\flux$ as variables using the \textit{E.coli} core model~\cite{orth2010reconstruction,palsson2015systems} and randomly chosen {stoichiometric} coefficients for the single constraint (Fig.~\ref{fig:fig2}). 
In this numerical calculation, metabolic pathways from exchangeable species to objective components are chosen as linear combinations of extreme pathways or elementary flux modes~\cite{schilling2000theory} for stoichiometry without objective reaction $o$ [Fig.~\ref{fig:fig2}(b)], although the above arguments do not depend on the specific choices of metabolic pathways (see {SM, Sec.~S3} for details). 
As shown in Fig.~\ref{fig:fig2}(a), the {linear} relation~\eqref{eq:Slutsky_Metabolism} {between metabolic responses} is indeed satisfied. Notably, {it is satisfied} regardless of the number and type of constraint(s) $\Const$, whereas the metabolic states themselves can sensitively depend on the concerned constraints and environmental conditions.

\textit{Relation between responses of reaction fluxes.---}
Although Eq.~\eqref{eq:Slutsky_Metabolism} generally holds for {arbitrary metabolic pathways}, it may be experimentally easier to manipulate a single metabolic reaction. 
Manipulation of a single reaction can affect multiple pathways because {they are} often tangled via a common reaction in {the metabolic network}. Thus, we should consider {the} contributions of multiple pathways. 
The simplest way for this is to sum up Eq.~\eqref{eq:Slutsky_Metabolism} for all the pathways that include the perturbed reaction $i$. 
However, to precisely conduct this summation, we need to know the whole stoichiometry matrix or metabolic network. 
Hence, another relation closed only for the reaction fluxes $\vflux$ is required for application without the need to know the details of the metabolic systems. 

To derive such a relation, we consider effective changes in the {stoichiometric} coefficients $S_{\alpha i}$ for reaction $i$ 
as metabolic inhibitions: e.g., inhibition of enzymes, administration of metabolite analogs, leakage of metabolites, 
and inefficiency in {the} allocation of some resource. 
We then obtain an equality on the optimized reaction fluxes $\hat{\vflux}$, formally similar to Eq.~\eqref{eq:Slutsky_Metabolism} (see {SM, Sec.~S4} for derivation):
\begin{eqnarray}\label{eq:Quantitative_Slutsky_v}
\frac{\partial \hat{v}_i(S, \Income)}{\partial q^\nu_i} = -\hat{v}_i(S, \Income)\frac{\partial \hat{v}_i(S, \Income)}{\partial I_\nu}
\end{eqnarray}
by defining the metabolic price $q^\nu_i$ of reaction $i$ in terms of $\nu$ as a function of $S$, instead of the metabolic price $p^\nu_\iprime$ of pathway $\iprime$ as a function of $K$, 
\begin{eqnarray}\label{eq:reaction_price}
q^\nu_i := \sum_{\alpha\in\Metabo\cup\Const} -S_{\alpha i} \left. \frac{\partial \hat{v}_i}{\partial I_\alpha} \middle/ \frac{\partial \hat{v}_i}{\partial I_\nu} \right..
\end{eqnarray}
The coefficient ${\left. \frac{\partial \hat{v}_i}{\partial I_\alpha} \middle/ \frac{\partial \hat{v}_i}{\partial I_\nu} \right.}\,{=:}\,c^\nu_\alpha(i)$ 
quantifies the number of units of component $\nu$ that can compensate for one unit of $\alpha$ in reaction $i$ and is experimentally measurable. 
For example, if $\nu$ is glucose and $\alpha$ is another metabolite such as an amino acid, $c^\nu_\alpha$ indicates how many units of glucose are required {to compensate} for one unit of the amino acid, similar to the ``glucose cost'' in previous studies~\cite{chen2022yeast}. 

For the linear response relation~\eqref{eq:Quantitative_Slutsky_v}, it is sufficient to calculate only the change in metabolic price (not the metabolic price itself), which depends on the type of manipulations {in concern}: (I) manipulations leading to the loss of a single component and (II) those leading to the loss of multiple components.

\begin{figure}[tb]
    \centering\includegraphics[width=\linewidth]{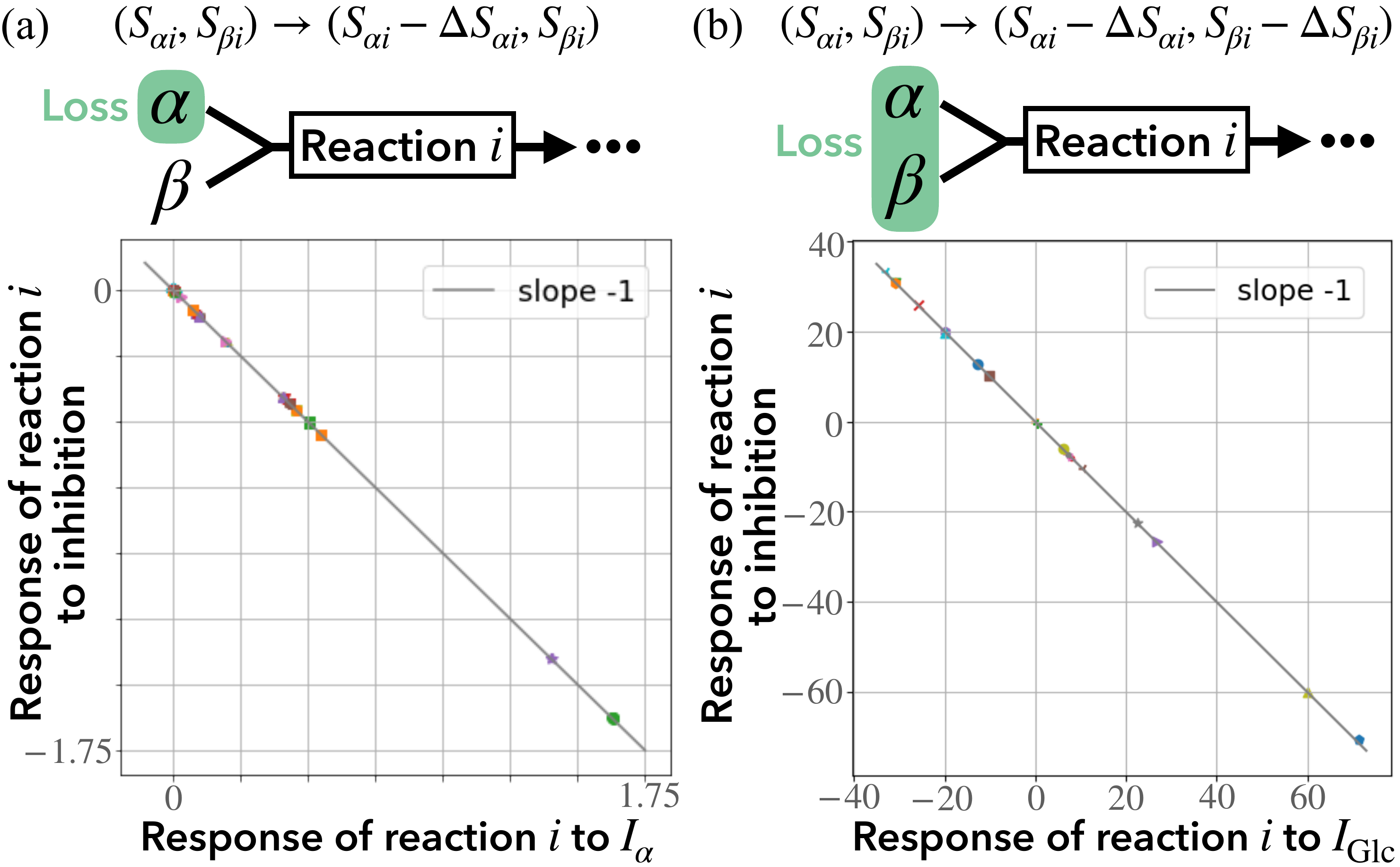}
    \caption{Responses of the optimized reaction fluxes $\hat{\vflux}$ against metabolic inhibitions of reaction $i$. 
    As the simplest example, a two-body reaction of components $\alpha$ and $\beta$ is illustrated in the upper panels. 
    (a) Metabolic inhibition of a single component
    ($\alpha$ in upper panel): case (I). 
    The horizontal axis shows the responses to the available amount of a constraint $\alpha\,(\in\Const)$, $\hat{v}_i \Delta \hat{v}_i(\Delta I_\alpha) / \Delta I_\alpha$, and the vertical axis does those to metabolic inhibitions, $\Delta \hat{v}_i(\Delta S_{\alpha i}) / \Delta q^\alpha_i$. 
    (b) Metabolic inhibition of multiple components 
    ($\alpha$ and $\beta$ in upper panel): case (II). Responses of the reaction flux $\hat{v}_i$ to effective changes in the input {stoichiometric} coefficients $\{S_{\mu i}\}_{\mu\in\Metabo}$ {for reaction $i$}, $\Delta \hat{v}_i(\{\Delta S_{\mu i}\}_{\mu\in\Metabo}) / \Delta q^\mathrm{Glc}_i$, are plotted against those to intake changes, $\hat{v}_i \Delta \hat{v}_i(\Delta I_\mathrm{Glc}) / \Delta I_\mathrm{Glc}$. 
    $I_\mathrm{Glc}\,{=}\,8.05\,\mathrm{[mmol/gDW/h]}$.  
    Each marker denotes a different reaction $i$. 
    }\label{fig:fig3}
\end{figure}

If experimental manipulation causes the loss of a single component $\alpha\,(\in\Metabo\cup\Const)$ in reaction $i$, $S_{\alpha i}$ effectively changes only for that $\alpha$ [Fig.~\ref{fig:fig3}(a)]. In such a case, the metabolic price change is just given by $\Delta q^\alpha_i\,{=}\,\Delta S_{\alpha i}$. 
An example of such experimental manipulations is the administration of an analog to a reactant of a multibody reaction: if $\alpha$ and $\beta$ react [see Fig.~\ref{fig:fig3}(a)], the metabolic analog of $\beta$ can produce incorrect metabolite(s) with $\alpha$, leading to the loss of $\alpha$, and thus, reaction $i$ requires more $\alpha$ to produce the same number of products, causing effective increases in the input {stoichiometric coefficient} $|S_{\alpha i}|$. Another example is the changes in the total capacity and effective stoichiometry for a constraint: for example, the mitochondrial volume capacity will work as such a constraint and can be genetically manipulated~\cite{malina2021adaptations,van2001modulating,raghevendran2006hap4}. Equation~\eqref{eq:Quantitative_Slutsky_v} for case (I) is numerically confirmed in Fig.~\ref{fig:fig3}(a). 

If metabolic inhibition of multiple reactant species of a reaction $i$ is simultaneously caused, the {stoichiometric} coefficients $S_{\mu i}$ for multiple reactants $\mu$ of reaction $i$ will effectively change [Fig.~\ref{fig:fig3}(b)]. Accordingly, the reaction price changes by $\Delta q^\nu_i\,{=}\,\sum_\mu \Delta S_{\mu i} c^\nu_\mu(i)$. In experiments, such cases would correspond to {the} inhibition of enzymes, leakage of the intermediate complex of reaction $i$, and so forth. 
Even in this case (II), the linear relation~\eqref{eq:Quantitative_Slutsky_v} is verified by numerically calculating the price changes of reaction $j$ defined in Eq.~\eqref{eq:reaction_price} 
{with the \textit{E.coli} core model including 77 reactions [Fig.~\ref{fig:fig3}(b)] as well as a larger-scale metabolic model including 931 reactions~\cite{reed2003expanded} (SM, Fig.~\ref{fig:Ec_iJR904}). 
Here, although} 
the precise calculation of the coefficients $c^\nu_\mu(i)$ requires information regarding not only the responses of $\hat{v}_i$ to $I_\nu$ but also those to $I_\mu$, they can be approximated in ways easier and independent of reaction $i$. For example, under extreme situations in which only the carbon sources limit the objective reaction, $c^\nu_\mu$ should be the ratios of the carbon numbers of species $\mu$ and $\nu$; alternatively, the simplest approximation could be just taking $c^\nu_\mu$ as unities. Even with these approximations, the relation~\eqref{eq:Quantitative_Slutsky_v} appears to hold well (SM, Fig.~\ref{fig:c_nu_approx}), and thus such approximations will be useful for qualitatively predicting whether metabolic inhibition promotes or suppresses the reaction of interest. 

Remarkably, our above argument does not depend on specific choices of objective reaction $o$, whereas we have utilized the biomass synthesis reaction as $o$ (Figs.~\ref{fig:fig2}~and~\ref{fig:fig3}). 
To highlight the independence of the relation~\eqref{eq:Quantitative_Slutsky_v} from cellular objective functions, we also numerically confirmed that it is satisfied even when objective reaction $o$ is set as a reaction for matter production, such as ethanol or ATP synthesis {(SM, Fig.~\ref{fig:SI_other_objective})}. These synthesis reactions are often considered as the objectives for metabolically engineered cells~\cite{stephanopoulos1998metabolic,schuetz2007systematic,gianchandani2008predicting}.

In the present study, we showed that the metabolic responses against resource availability and those against metabolic inhibitions are negatively proportional. The quantitative relations we found should be universally satisfied with arbitrary reaction networks, constraints, and objective functions of cells. In particular, although the predicted optimal metabolic states can drastically depend on the assumed objective function, the relations of the responses should be always satisfied independent of it {(see also SM, Fig.~\ref{fig:SI_other_objective})}. 
Even though we can never know the true objective function of cells, we can still predict the metabolic responses. 

{
In the linear relations, the metabolic responses against different perturbations are linked because both are determined from the identical objective function and constraints (see also Fig.~\ref{fig:fig1}). It is similar to the linear response theories in statistical mechanics: they are derived from the fact that different thermodynamic quantities are given as the derivatives of an identical thermodynamic potential~\cite{LANDAU1980333}. 
Note here that the thermodynamic potential works as an objective function: e.g., entropy is maximized at the thermal equilibrium. 
} 

The independence from cellular objective functions is derived from the microeconomic formulation for metabolic regulation and application of the Slutsky equation in economics. Note that the Slutsky equation basically requires detailed information regarding the objective functions (utility functions in economics) because it includes a term for the so-called substitution effect that quantifies the substitutability of goods and depends on the objective functions (see also {SM, Sec.~S4}). However, the term disappears when applied to metabolism {because the law of mass conservation implies the non-substitutability of reactants}. 

{
Although the linear relations~(\ref{eq:Slutsky_Metabolism}-\ref{eq:Quantitative_Slutsky_v}) are general due to the generality of the law of mass conservation, there are also some limitations. First, since our results rely on the assumption of optimal metabolic regulation, they will not hold in suboptimal metabolic responses; conversely, any observed deviation from the relations~(\ref{eq:Slutsky_Metabolism}-\ref{eq:Quantitative_Slutsky_v}) will indicate the suboptimality in the regulation of the real metabolic system in question. 
Second, our linear relations work only for continuous metabolic responses. 
Third, the approximation of coefficients $c_{\mu}^{\nu}$ in Eqs.~(\ref{eq:Quantitative_Slutsky_v}-\ref{eq:reaction_price}) could be prohibitive when the coefficients become negative, e.g., in the case the fluxes from different nutrient sources must be balanced for a metabolic reaction of interest and an increase in one source promotes the reaction while an increase in another source inhibits it.} 

Because our results are valid regardless of how abstract the concerned model is, from coarse-grained toy models to genome-scale metabolic networks, they would be important both for quantitative predictions and for discovering qualitatively novel phenomena. The Warburg effect {or overflow metabolism} is a prominent example of the latter. 
In the Warburg effect, {as} the amount of the carbon source taken up by a cell increases, the cell decreases the flux of the respiration pathway {and utilize fermentation or aerobic glycolysis instead}~\cite{Heiden-Warburg,OMbook}. From the relation~\eqref{eq:Slutsky_Metabolism}, one can immediately predict that the inhibition of respiration (e.g., administration of uncouplers of respiration~\cite{WeakAcid2}) will counterintuitively increase the respiration flux. 
Such an increase in the respiration flux was observed in a coarse-grained model, which was termed the drug-induced reverse Warburg effect~\cite{yamagishi2021microeconomics}. 
This phenomenon has been indeed reported in several published experiments~\cite{WeakAcid,WeakAcid2,WeakAcid3,Therapies-induced-reversed-Warburg}. 

{Likewise, for controlling cellular metabolic states, e.g., for metabolic engineering and medicine, some counterintuitive manipulations can promote pathway or reaction fluxes. Although metabolic inefficiency is considered to suppress the flux in general, when an increase in the intake of a substrate suppresses a pathway or reaction flux, making the metabolic pathway or reaction less efficient will counterintuitively promote the flux (see also SM, Sec.~S2 and Fig.~\ref{fig:SI_toy_model} for an example of coarse-grained models).} 
In experimental application, the intake or total capacity can be altered by shifts in environmental conditions, genetic manipulations, and so forth. Changes in the metabolic price can be also implemented in various ways: e.g., administration of a metabolite analog, leakage of a metabolite, {addition of an alternative pathway or reaction through metabolic engineering manipulations,}
and inhibition of some enzyme that will lead to {the} accumulation of the reactants and possibly promote their excretion or conversion to other chemicals. They cause a loss of reactants, and thus, the corresponding reaction(s) require more metabolites to produce the same number of products. 

The relations~(\ref{eq:Slutsky_Metabolism}-\ref{eq:Quantitative_Slutsky_v}) allow us to predict the responses of {an arbitrary} reaction or pathway flux to metabolic inhibitions only by measuring its fluxes in several different nutrient conditions, and vice versa. {The} predictions do not require detailed information regarding the concerned intracellular reaction networks, and they are valid even when the precise estimation of effective changes in the {stoichiometric} coefficients is difficult, at least qualitatively (SM, Fig.~\ref{fig:c_nu_approx}). 
Therefore, they will be useful as {quantitative and qualitative guidelines} to operate the metabolic states toward the desirable directions in various fields such as microbiology, metabolic engineering, and medicine. 

{
The associated Python code and data to reproduce figures in this work are available~\footnote{\url{https://github.com/JFYamagishi/yamagishi-hatakeyama-2023}}.} 

{
% \begin{acknowledgments}
We would like to acknowledge Wolfram Liebermeister, Chikara Furusawa, Yasushi Okada, Kunihiko Kaneko, and Takuma \={O}nishi for helpful discussions and useful comments. 
This work was partially supported by JSPS KAKENHI Grant Numbers JP21J22920 and JP21K15048. 
% \end{acknowledgments}
}

% % \bibliographystyle{apsrev4-2}
% \bibliographystyle{savetrees}
% \bibliographystyle{ieeetr}
% \bibliographystyle{unsrt}
\bibliographystyle{elsarticle-num}

\bibliography{bibliography_2023-07.bib}

\clearpage
\onecolumngrid
% \appendix
\setcounter{figure}{0}
\renewcommand{\thefigure}{S\arabic{figure}}
\setcounter{equation}{0}
\renewcommand{\theequation}{S\arabic{equation}}
\setcounter{section}{0}
\renewcommand{\thesection}{S\arabic{section}}
\renewcommand{\thesubsection}{S\arabic{subsection}}

\newcommand{\nPath}{|\mathcal{P}|}
\newcommand{\nReac}{|\mathcal{R}|}

\begin{center}
    {\large\textbf{Supplemental Material}}
\end{center}

\section{Equivalence between constraint-based modeling (CBM) and microeconomic formulations} 
\noindent\textbf{CBM formulation: optimization problems with reaction fluxes $\vflux$ as variables.} 
In the framework of CBM in systems biology, intracellular metabolic regulation is formulated as linear programming (LP) problems in which the variables are the fluxes $\vflux$ of reactions. 
As discussed below, LP problems in CBM are generally equivalent to optimization problems in the microeconomic theory of consumer choice. 

By breaking down each reversible reaction into two irreversible reactions (i.e., its forward and backward components), a non-negative $\nReac$-dimensional vector $\vflux:=\{v_i\}_{i\in\Reac}$ represents the fluxes of all reactions. 
{Then, by} assuming the stationarity of the intracellular concentrations of non-exchangeable species {$\mu\;(\in\Metabo\backslash\mathcal{E})$}, {a} general formulation of CBM~\cite{palsson2015systems,klipp2016systems} is given as follows: 
\begin{eqnarray} \label{eq:CBM}
&\max_{\vflux\geq {\bf 0}}&\;v_o \;\; \mathrm{s.t.} \nonumber \\
&&\sum_{j\in\Reac}S_{\mu j}v_j=0\;\;(\mu\in\Metabo\backslash\mathcal{E})\label{eq:Balanced} \\
&&\sum_{j\in\Reac}S_{\alpha j}v_j + I_\alpha \geq 0\;\;(\alpha\in\mathcal{E}\cup\Const) \label{eq:Leakage}
\end{eqnarray}
Because $\sum_{j\in\Reac}S_{\mu j}v_j$ is equal to the excess production of species $\mu$, Eq.~(\ref{eq:Balanced}) represents that the production and degradation of internal metabolites must be balanced. 
With respect to Eq.~(\ref{eq:Leakage}), if $\alpha$ is a species ($\alpha\in\Excess\subset\Metabo$), the corresponding inequality, $\sum_{j\in\Reac}S_{\alpha j}v_j + I_\alpha \geq 0$, represents that {exchangeable species $\alpha$ with intake $I_\alpha>0$ are taken in and species $\alpha$ with efflux $I_\alpha<0$} are (forcibly) leaked or degraded; in contrast, if $\alpha$ is a constraint ($\alpha\in\Const$), it represents {a non-stoichiometric constraint,} e.g., allocation of some limited resource such as proteins~\cite{scott2011bacterial,OM}, intracellular space~\cite{vazquez2010catabolic,OMbook}, membrane surface~\cite{memRealEstate2}, and Gibbs energy dissipation~\cite{Heinemann-Gibbs2019}. {Stoichiometric coefficient} $S_{\alpha i}$ for {some constraint} $\alpha\,(\in\Const)$ is typically negative or zero but can also be positive. 
Note that $\Const$ can also include other constraints like the upper and lower bounds of the flux of reaction $i$.

\noindent\textbf{Optimization problems with pathway fluxes $\f$ as variables.} 
The usual CBM formulation (\ref{eq:Balanced}-\ref{eq:Leakage}) with reaction fluxes $\vflux$ as variables is equivalent to another LP problem with pathway fluxes $\f$ as variables (see also Fig.~\ref{fig:fig1} in the main text). 

Here, a metabolic pathway is a linked sequence of reactions. 
Pathway fluxes $\f$ are related to reaction fluxes as $\vflux=P\f$, with a \textit{pathway matrix} $P:=\{P_{i\iprime}\geq0\;|\;i\in\Reac,\iprime\in\Path\}$ which represents that pathway $iprime$ comprises $P_{i\iprime}$ units of reaction $i$. The stoichiometry matrix $K$ for pathways expresses the metabolic pathways in ``species space'' and is related to the stoichiometry matrix $S$ for reactions with $K=SP$. 

When $P$ is taken as (linear combinations of) elementary flux modes (EFMs; i.e., extreme rays of the flux cone), Eq.~(\ref{eq:Balanced}) is {autonomously} satisfied ({and matrix $K:=SP$ is then related to so-called elementary conversion modes~\cite{clement2021unlocking}}). Then, the LP problem (\ref{eq:Balanced}-\ref{eq:Leakage}) for CBM can be rewritten into another LP problem with pathway fluxes $\f$ as variables:
\begin{eqnarray} \label{eq:n_goods-m_objectives_Supp}
\max_{\Lambda,\f\geq {\bf 0}}\; 
\Lambda
\quad\mathrm{s.t.} \quad I_\alpha + \sum_{j^\prime\in\Path}K_{\alpha j^\prime}f_{j^\prime} \geq -S_{\alpha o}\Lambda  \;\;(\alpha\in\mathcal{E}\cup\Const) 
\end{eqnarray}
where the components required for objective reaction $o\,(\in\Reac)$ are termed as objective components $\Obj := \{\alpha\in\Metabo\cup\Const\;|\;S_{\alpha o}<0\}$.

\noindent\textbf{Derivation of the microeconomic formulation.} 
The min function for the objective function, $\Lambda:=\min(A,B)$, can be replaced by constraints of $\Lambda\leq A$ and $\Lambda\leq B$, and vice versa. 
Therefore, the optimization problem~(\ref{eq:n_goods-m_objectives_Supp}) is equivalent to the optimization of the Leontief utility function [Eq.~(\ref{eq:growth}) in the main text]---the minimum of multiple ``complementary'' objectives---under the constraints of Eq.~(\ref{eq:n_goods-m_objectives}) in the main text. 
In contrast, the microeconomic formulation of optimization problems (Eqs.~(1-2) in the main text) can be converted to LP problems in the form of Eq.~(\ref{eq:n_goods-m_objectives_Supp}), and thus, they are equivalent.

{ % \color{red}
\section{A simple example of microeconomic formulation of metabolic regulation}
As a concrete example, we here introduce a simple, analytically-solvable metabolic model (see also Fig.~\ref{fig:SI_toy_model}). It can be the simplest model of a metabolic system in which multiple pathways can produce a common metabolite from a single resource, e.g., the co-utilization of respiration and fermentation or that of Embden–Meyerhoff–Parnass (EMP) and Entner–Doudoroff (ED) glycolytic pathways~\cite{EMP/ED}. 

This model consists of $2$ metabolites, $\Metabo=\{m_1,m_2\}$, and $1$ constraint $\Const=\{\rho\}$, as well as $3$ reactions, $\Reac=\{A,B$, objective$\}$. 
Note that $\rho$ can be any kind of limited resource, e.g., such as the intracellular volume or solvent capacity~\cite{vazquez2010catabolic,OMbook}, the total amount of proteins~\cite{OM}, and the total area of membrane surfaces~\cite{memRealEstate2}. 
We assume that $m_1$ is, out of $\Metabo$, the only exchangeable species, i.e., $\Excess=\{m_1\}$, reactions $A,B$ both produce $m_2$ by consuming $m_1$ and $\rho$, and the objective reaction $o$ requires $m_2$ and $\rho$. 
The stoichiometry matrix, $S\in\mathbb{R}^{(\Metabo\cup\Const)\times\Reac}$, is then given as 
\begin{eqnarray*}
\begin{split}
  S = \begin{pmatrix}
    -p^{m_1}_A & -p^{m_1}_B & 0 \\
    S_{m_2A} & S_{m_2B} & -1 \\
    -p^\rho_{A} & -p^\rho_{B} & - 1
    \end{pmatrix},
\end{split}
\end{eqnarray*}
with positive parameters (i.e., ``price'' of metabolic pathways) $p^\alpha_i$ $(\alpha=m_1,\rho;\; i=A,B)$. For simplicity, the stoichiometric coefficients are normalized as $|S_{m_2 o}| = |S_{\rho o}| = 1$, whereas it does not lose its generality. 
With this matrix, the LP problem for CBM with reaction fluxes $\vflux$ as variables is represented as 
\begin{eqnarray*} \label{eq:LP_v}
\max_{\vflux\geq \textbf{0}}\; v_\mathrm{BM}
\quad
\mathrm{s.t.}
\quad
&&\sum_{j}S_{\mu j}v_j + I_\mu \geq 0 \quad(\mu=m_1,\rho),\\
&&\sum_{j}S_{m_2, j}v_j = 0.
\end{eqnarray*}

In this simple model, reactions $A,B$ are metabolic pathways as they are, since they connect the nutritional species $m_1$ to objective component $m_2$. Then, we do not distinguish the fluxes of reactions $\vflux$ and those of metabolic pathways $\flux$. 
Thus, the microeconomic formulation of the above LP problem is the maximization of 
\begin{eqnarray*}\label{eq:LP_ve}
% \max_{\f\geq 0}\; 
\Lambda(\flux):= 
\min\left[
S_{m_2A}f_A + S_{m_2B}f_B
,\;
I_\rho - p^\rho_Af_A - p^\rho_Bf_B
\right]
\quad\mathrm{s.t.}\quad
p^{m_1}_Af_{A} + p^{m_1}_Bf_{B} \leq I_{m_1},
\end{eqnarray*} 
and the relations~(\ref{eq:Slutsky_Metabolism})~and~(\ref{eq:Quantitative_Slutsky_v}) in the main text are equivalent in this model: 
\begin{eqnarray}\label{eq:Slutsky_SM}
\frac{\partial \hat{f}_{i}}{\partial p^{m_1}_j} = -\hat{f}_{j}\frac{\partial \hat{f}_i}{\partial I_{m_1}}. \quad (i,j=A,B)
\end{eqnarray}

We can analytically obtain the optimized solutions for this model, from which we can directly confirm that the relation~(\ref{eq:Slutsky_SM}) is indeed satisfied as follows. 
Since the solutions depend on the relative values of the parameters, we analyze the behavior of the model here by dividing it into two cases: whether some reaction flux is suppressed by the increase in nutrient intake $I_{m_1}$ or not.

\noindent\textbf{When no reaction is suppressed by increasing the nutrient intake.} 
Without any trade-off between reactions $A,B$, only the more efficient reaction is  used in the optimized solutions. In such case, no reaction flux is suppressed by increasing the nutrient intake and metabolic inhibition of a reaction always causes its suppression. 

For example, when reaction $A$ is more efficient, the optimal flux $\hat{\flux}$ is just given by 
\begin{eqnarray}\label{eq:solution_evident}
\left(\hat{f}_{A},\hat{f}_{B}\right) = \left(
\min\left[
\frac{I_{m_1}}{p^{m_1}_A}
,
\frac{I_\rho}{p^\rho_A}
\right],
0
\right),
\end{eqnarray}
and the flux $\hat{f}_{A}$ is promoted or sustained by increasing $I_{m_1}$. 
Also, from Eq.~(\ref{eq:solution_evident}), when $I_{m_1}/p^{m_1}_A \leq I_\rho/p^\rho_A$ holds, the metabolic responses are given as 
\begin{eqnarray*}
\frac{\partial \hat{f}_A}{\partial I_{m_1}} = \frac{1}{p^{m_1}_A},\quad 
\frac{\partial \hat{f}_A}{\partial p^{m_1}_A} = -\frac{I_{m_1}}{(p^{m_1}_A)^2},\quad 
\frac{\partial \hat{f}_A}{\partial p^{m_1}_A} = -\hat{f}_A \frac{\partial \hat{f}_A}{\partial I_{m_1}} 
,
\end{eqnarray*}
and 
\begin{eqnarray*}
\hat{f}_A\frac{\partial \hat{f}_B}{\partial I_{m_1}} = \frac{\partial \hat{f}_B}{\partial p^{m_1}_A} = 0,\quad
\hat{f}_B\frac{\partial \hat{f}_i}{\partial I_{m_1}} = \frac{\partial \hat{f}_i}{\partial p^{m_1}_B} = 0\;\;(i=A,B),
\end{eqnarray*}
and thus the relation~(\ref{eq:Slutsky_SM}) holds. In the case $I_{m_1}/p^{m_1}_A > I_\rho/p^\rho_A$, the optimal fluxes are independent of the value of $I_{m_1}$ and the relation~(\ref{eq:Slutsky_SM}) is evidently satisfied as 
\begin{eqnarray*}
\hat{f}_j\frac{\partial \hat{f}_i}{\partial I_{m_1}} = \frac{\partial \hat{f}_i}{\partial p^{m_1}_j} = 0\;\;(i,j=A,B).
\end{eqnarray*}

\noindent\textbf{When reaction $A$ is suppressed by increasing the nutrient intake.} 
With some trade-off between reactions $A,B$, the situation is non-trivial: 
the co-utilization of different pathways $A,B$ and switching between them can be observed. In particular, we here assume a trade-off $s_{m_2A} > s_{m_2B}$ and $p^\rho_A > p^\rho_B$, i.e., reaction $A$ is more efficient in producing $m_2$ but requires more amount of constraint $\rho$ than the alternative reaction $B$. When the trade-off matters, the optimized flux of reaction $A$ is suppressed by increasing nutrient intake $I_{m_1}$ and it is promoted by its metabolic inefficiency. 

\begin{figure}[tb]
    \centering \includegraphics[width = 0.8\linewidth]{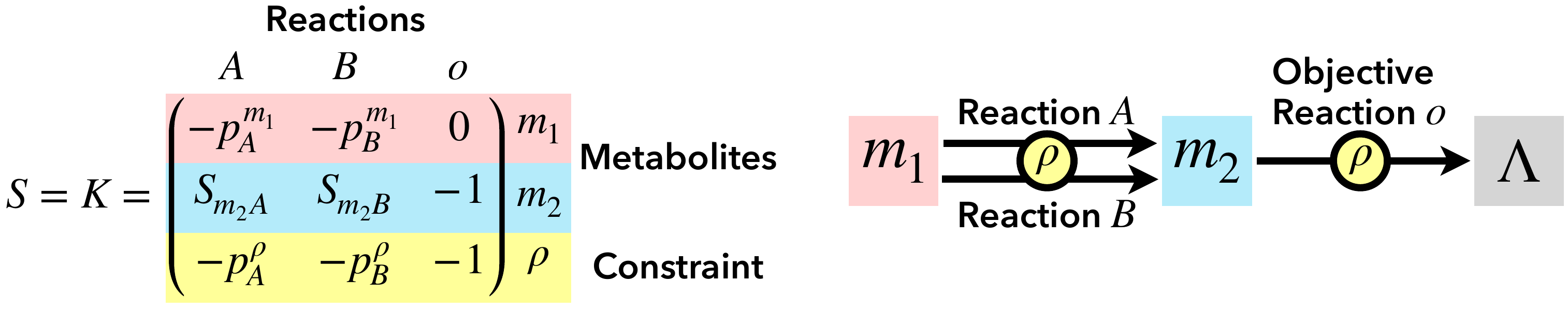}
    \caption{
        {A simple example of microeconomic formulation of metabolic regulation.} 
    }\label{fig:SI_toy_model}
\end{figure}

We also assume that $p^{m_1}_A$ is not so large compared to $p^{m_1}_B$ that $p^{m_1}_B B_0 > p^{m_1}_A A_0$ holds, where $ A_0 := I_\rho / (S_{m_2A} + p^\rho_A)$ and $ B_0 := I_\rho / (S_{m_2B} + p^\rho_B)$. The optimal fluxes $\hat{\flux}$ is then calculated as
\begin{eqnarray*} \label{eq:p>1}
\left(\hat{f}_{A},\hat{f}_{B}\right)
=\begin{cases} \left(\frac{I_{m_1}}{p^{m_1}_A}, 0 \right)
\quad \mathrm{if}\;I_{m_1}\leq p^{m_1}_A A_0\\
\left(
\frac{p^{m_1}_B B_0-I_{m_1}}{p^{m_1}_BB_0 - p^{m_1}_AA_0}A_0
,
\frac{I_{m_1} - p^{m_1}_A A_0}{p^{m_1}_BB_0 - p^{m_1}_A A_0}B_0
\right)
\quad \mathrm{if}\; p^{m_1}_B B_0\geq I_{m_1}\geq p^{m_1}_A A_0\\
\left(0, B_0\right)
\quad \mathrm{if}\;I_{m_1}\geq p^{m_1}_B B_0
\end{cases}
\end{eqnarray*}
From the dependence of this solution on $I_{m_1}$ and $p^{m_1}_i$, we can immediately confirm that the relation~(\ref{eq:Slutsky_SM}) is satisfied. 

In the case of $I_{m_1}\leq p^{m_1}_A A_0$ or $I_{m_1} \geq p^{m_1}_B B_0$, either reaction is used and thus the relation~(\ref{eq:Slutsky_SM}) is evidently satisfied as in the above case. 

In contrast, with the medium value of the intake $I_{m_1}$ of metabolite $m_1$ that satisfies $p^{m_1}_B B_0\geq I_{m_1}\geq p^{m_1}_A A_0$, the metabolic responses against $ I_{m_1}$ are: 
\begin{eqnarray*} 
\frac{\partial \hat{f}_A}{\partial I_{m_1}} &=& 
- \frac{A_0}{p^{m_1}_BB_0 - p^{m_1}_AA_0} 
,\quad 
\frac{\partial \hat{f}_B}{\partial I_{m_1}} = 
\frac{B_0}{p^{m_1}_BB_0 - p^{m_1}_A A_0}.
\end{eqnarray*}
Note here that, as $p^{m_1}_B B_0 > p^{m_1}_A A_0$ holds from the assumption, 
\begin{eqnarray*}
\frac{\partial \hat{f}_A}{\partial I_{m_1}}<0,\;\frac{\partial \hat{f}_B}{\partial I_{m_1}}>0.
\end{eqnarray*}
Thus, we can immediately obtain the qualitative prediction that the inhibition of reaction/pathway $A$ promotes its own flux $\hat{f}_A$ as well as the quantitative prediction~(\ref{eq:Slutsky_SM}). Since the price responses are 
\begin{eqnarray*} 
\frac{\partial \hat{f}_A}{\partial p^{m_1}_A} = 
\frac{p^{m_1}_B B_0-I_{m_1}}{(p^{m_1}_BB_0 - p^{m_1}_AA_0)^2}(A_0)^2
= -\hat{f}_{A}\frac{\partial \hat{f}_A}{\partial I_{m_1}}
,&\quad&
\frac{\partial \hat{f}_A}{\partial p^{m_1}_B} = 
\frac{I_{m_1} - p^{m_1}_AA_0}{(p^{m_1}_BB_0 - p^{m_1}_AA_0)^2}A_0B_0
= -\hat{f}_{B}\frac{\partial \hat{f}_A}{\partial I_{m_1}}
,\\
\frac{\partial \hat{f}_B}{\partial p^{m_1}_A} = 
-\frac{p^{m_1}_B B_0-I_{m_1}}{(p^{m_1}_BB_0 - p^{m_1}_A A_0)^2}A_0B_0
= -\hat{f}_{A}\frac{\partial \hat{f}_B}{\partial I_{m_1}}
,&\quad&
\frac{\partial \hat{f}_B}{\partial p^{m_1}_B} = 
-\frac{I_{m_1} - p^{m_1}_A A_0}{(p^{m_1}_BB_0 - p^{m_1}_A A_0)^2}(B_0)^2
= -\hat{f}_B\frac{\partial \hat{f}_B}{\partial I_{m_1}}
,
\end{eqnarray*}
the linear relation~(\ref{eq:Slutsky_SM}) is indeed satisfied. 

From the biological perspective, by regarding reactions/pathways $A$ and $B$ as respiration and fermentation pathways, respectively, the above model can be interpreted as a minimal model for overflow metabolism or the Warburg effect~\cite{OMbook}. In one interpretation, $m_1$, $m_2$, $\rho$, and $o$ can be regarded as glucose, ATP, the solvent capacity, and biomass synthesis reaction, respectively. Then, the trade-off and the suppression of respiration against the increase in carbon intake are consistent with the empirical observations. 
Using our linear relation, we can quantitatively predict the stimulation of the respiration flux (i.e., the drug-induced reverse Warburg effect) by the administration of uncouplers of respiration. 
}

\section{Details of numerical experiments}\label{sec:SI_numerical_calc}
In our numerical simulations, stoichiometric coefficients $S_{\alpha i}$ for species $\alpha\in\Metabo$ are given by the \textit{E.coli} core model~\cite{orth2010reconstruction}. 
$S_{\alpha i}$'s for constraints $\alpha\in\Const$ are randomly chosen from an interval $[-1,0]$ to show that the results do not depend on such details; here, as a natural assumption, we set $S_{\alpha i}$ {to be identical for the pair of reactions split from the same reversible reaction.} 

In most numerical simulations, the objective reaction $o$ is chosen as the biomass synthesis reaction, whereas our results do not depend on the choice of the objective reactions: e.g., synthesis reaction of ethanol or ATP (see also Fig.~\ref{fig:SI_other_objective}). 

\begin{figure}[tbh]
    \centering\includegraphics[width=0.4\linewidth]{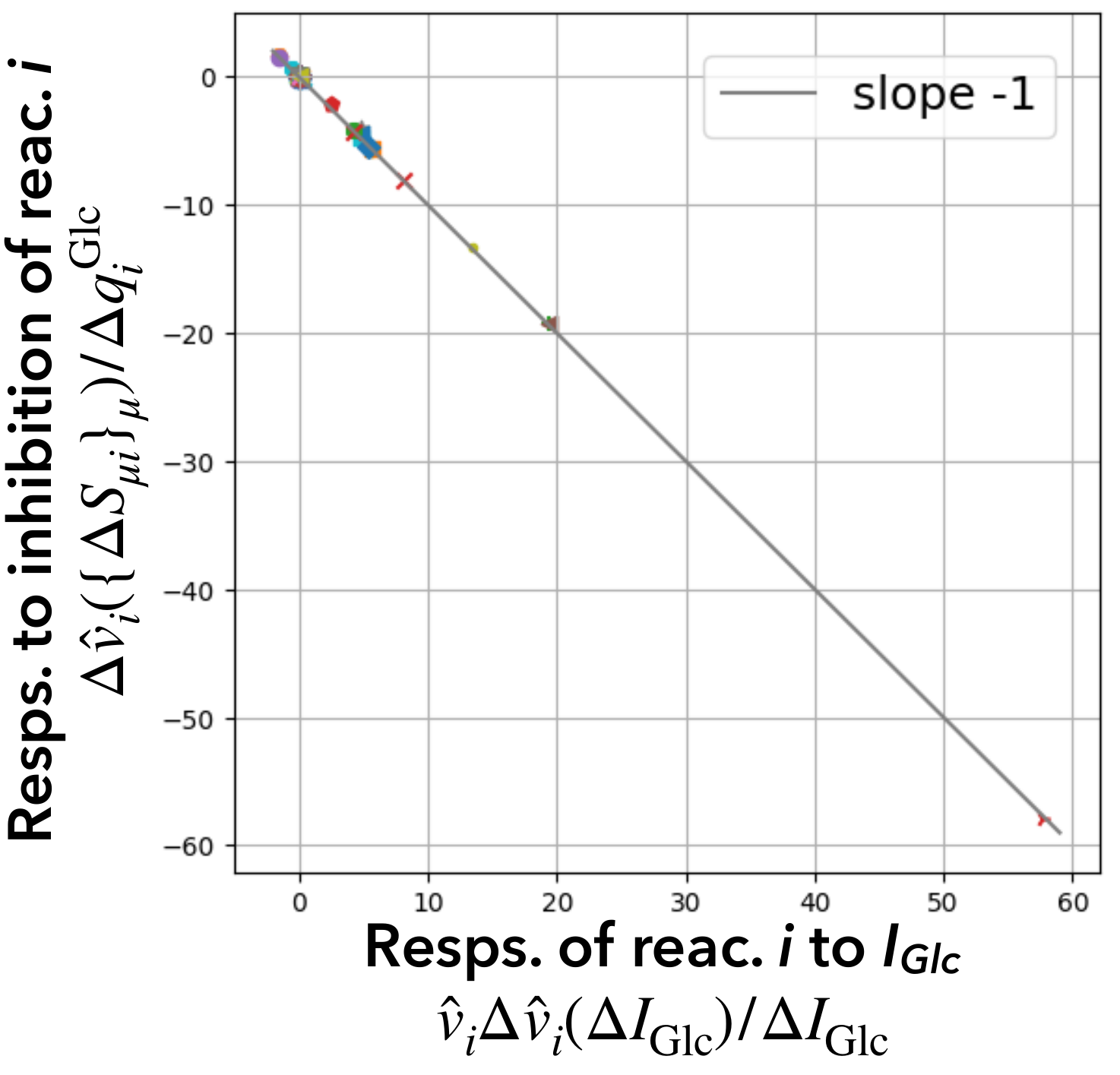} 
    \caption{
    {
    Numerical calculation with the genome-scale \textit{E.coli} iJR904 model~\cite{reed2003expanded} for responses of the optimized reaction fluxes $\hat{\vflux}$. 
    The metabolic responses to effective changes in the input stoichiometric coefficients $\{S_{\mu i}\}_{\mu\in\Metabo}$, $\Delta \hat{v}_i(\{\Delta S_{\mu i}\}_{\mu\in\Metabo}) / \Delta q^\mathrm{Glc}_i$, are plotted against those to intake changes, $\hat{v}_i \Delta \hat{v}_i(\Delta I_\mathrm{Glc}) / \Delta I_\mathrm{Glc}$. 
    $I_\mathrm{Glc}\,{=}\,8.075\,\mathrm{[mmol/gDW/h]}$.  
    Each marker denotes a different reaction $i$. 
    }
    }\label{fig:Ec_iJR904}
\end{figure}

\begin{figure*}[bht]
	\centering \includegraphics[clip, width=0.995\columnwidth]{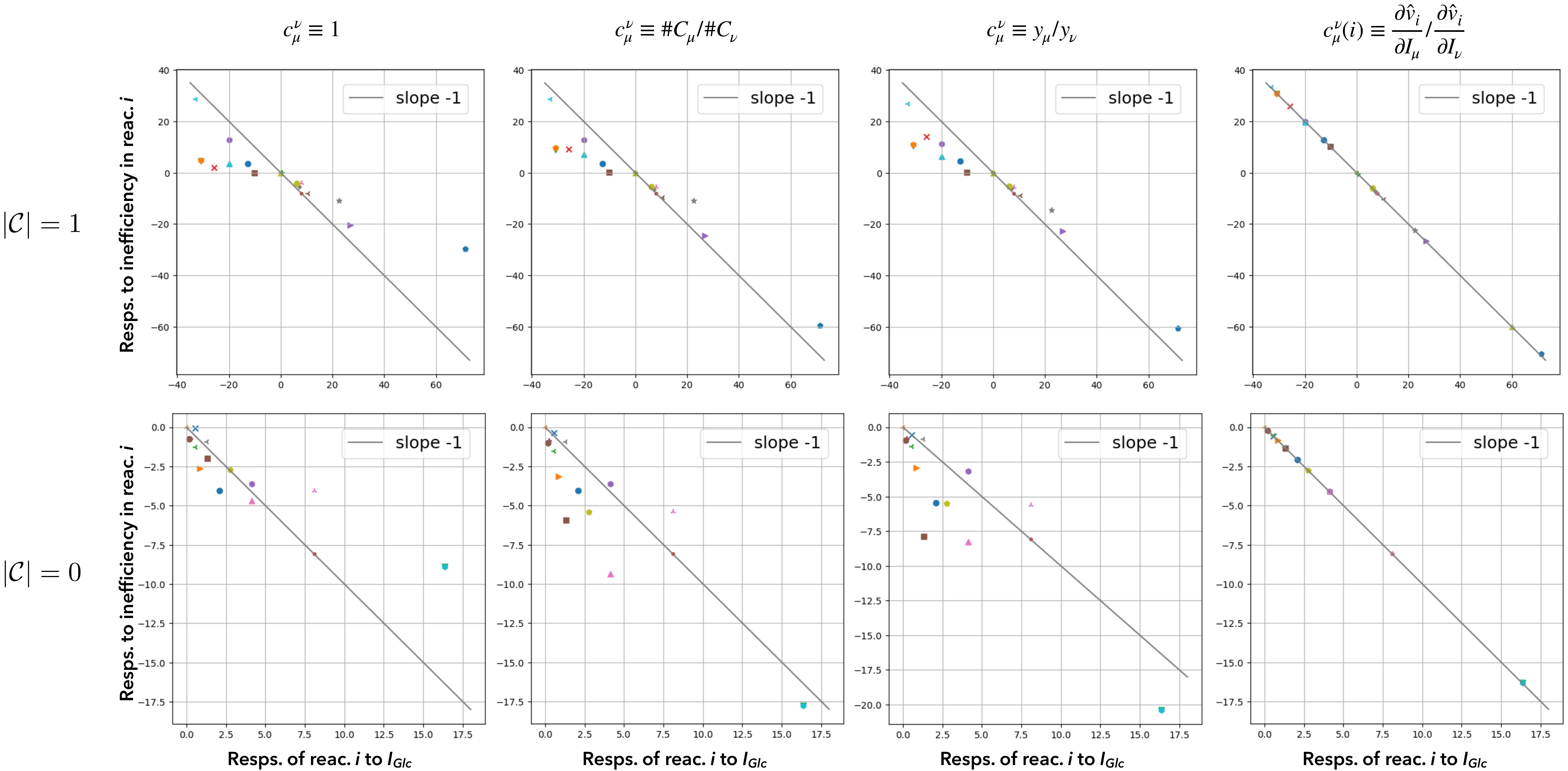}
    \caption{
    The metabolic responses with several approximated estimations of $c^\nu_\mu$ are plotted. 
    (Top) With a constraint (i.e., the number of elements in $\Const$, {$|\Const|$,} is 1). (Bottom) Without constraints (i.e., {$|\Const|$} is 0). 
    From the left, $c^\nu_\mu$ is approximated as $1$ independent of $\mu$, $c^\nu_\mu$ is approximated as the ratios of the carbon numbers of species $\mu$ and $\nu$, $c^\nu_\mu$ is approximated as $\frac{\partial \Lambda}{\partial I_\mu}/\frac{\partial \Lambda}{\partial I_\nu}$ (that equals to the ratio of the shadow price~\cite{reznik2013flux} of species $\mu$ and $\nu$, $y_\mu/y_\nu$), and Eq.~(\ref{eq:reaction_price}) in the main text (without approximation). 
    }\label{fig:c_nu_approx}
\end{figure*}

\begin{figure}[bth]
    \centering \includegraphics[width = 0.65\linewidth]{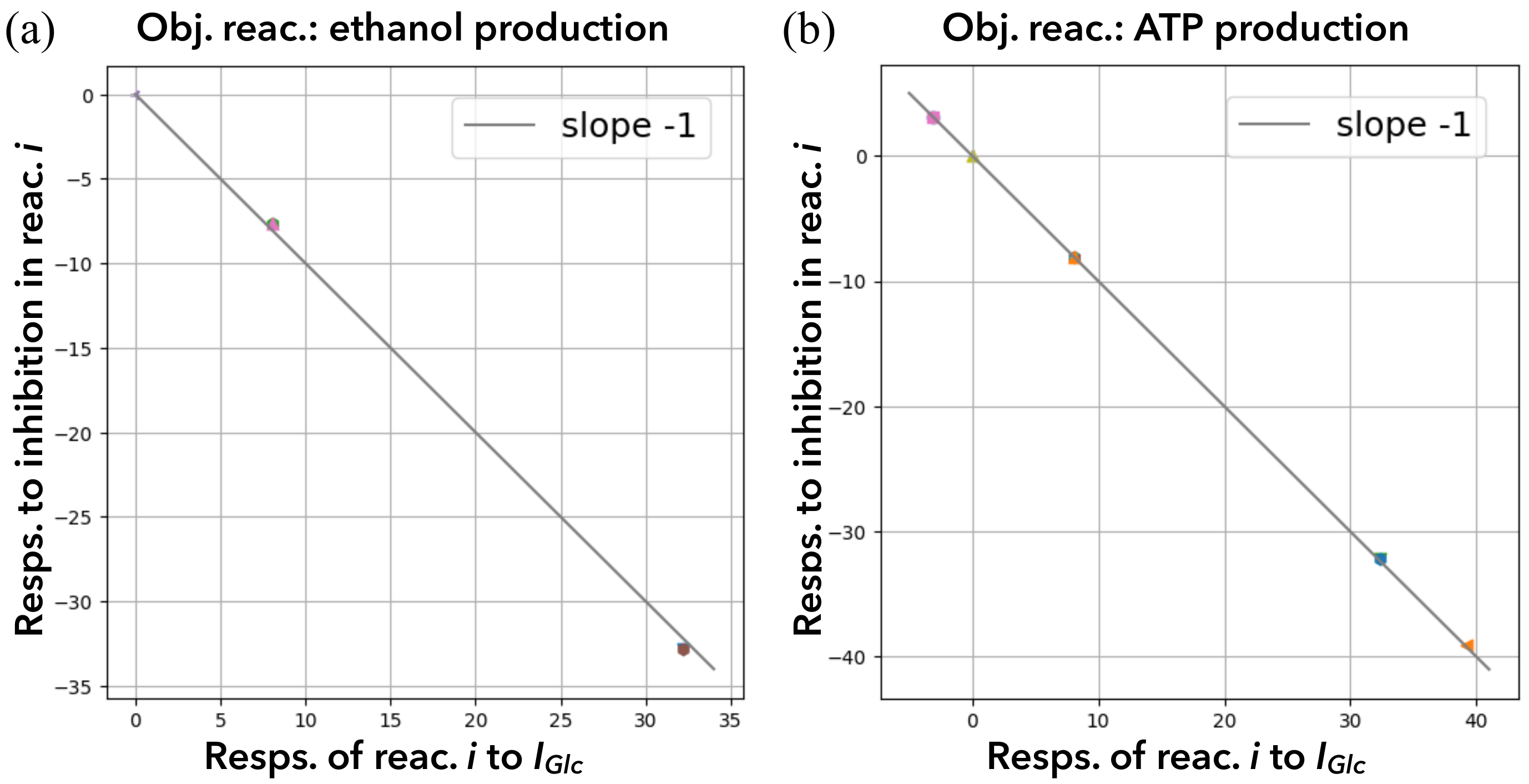}
    \caption{
    The metabolic responses with matter production as the objective reaction. $|\Const|= 0$. (A) Ethanol production (acetaldehyde dehydrogenase; abbreviated as ``ADHEr'' in \textit{E. coli} core model). (B) ATP production (ATP maintenance requirement; abbreviated as ``ATPM'' in \textit{E. coli} core model).
    }\label{fig:SI_other_objective}
\end{figure}

We plotted only reactions with continuous price responses in Figs.~2~and~3 in the main text and {Figs.~\ref{fig:Ec_iJR904}-\ref{fig:SI_other_objective}.} 
In our numerical calculations, the response to price change is considered to be discontinuous if decreasing $\Delta p$ changes the reaction flux discontinuously at $\Delta p\to0$.

\noindent\textbf{On the uniqueness of the solutions.} 
For simplicity, the arguments in the main text are based on the premise of the uniqueness of the solution. 
This assumption indeed holds true with suitable penalty terms explained below. Furthermore, it must be biologically natural because we considered the intracellular metabolic responses around a steady state here. 

An LP can sometimes have multiple solutions. If an LP has multiple solutions because of the existence of ineffective inequalities, it is natural to define the unique solution as the solution with the least consumption rates of species: i.e., slightly modify the objective function as $v_\mathrm{BM}-\epsilon\sum_{\alpha\in\Metabo\cup\Const}\left(\sum_{i\in\Reac}-S_{\alpha i}v_i\right)$ with small $\epsilon>0$ (here, note that $\sum_{i\in\Reac}-S_{\alpha i}v_i$ is equal to the consumption rate of $\alpha$, while $I_\alpha$ is its upper bound). 
In addition, for the pair of two irreversible reactions formed by separating each reversible reaction, the smaller of the two must be subtracted and set to zero for the uniqueness of the solution by adding a penalty $-\epsilon\sum_iv_i$.
Throughout the numerical calculations in the main text, $\epsilon$ is basically set to $\lesssim 10^{-5}$, whereas the results do not depend on $\epsilon$ as long as $\epsilon$ is sufficiently small.

\noindent\textbf{Leakage of intermediate species.} 
To calculate the reaction price $q^\nu_i$ (Eq.~(\ref{eq:reaction_price}) in the main text), we must calculate the responses of $\hat{v}_i$ against changes in $I_\mu$ for each reactant $\mu$ of reaction $i$. For calculating the responses to $I_\mu$ of each $\mu$, we assumed that species $\mu$ is exchangeable. Such changes do not alter the optimized solution $\hat{\vflux}$ in the numerical calculations we conducted.

\noindent\textbf{Choice of metabolic pathways in Fig.~2 in the main text.} 
In this study, we consider (linear sums of) EFMs or extreme pathways for the stoichiometry without objective reaction $o$ as the metabolic pathways. 
That is, we consider metabolic pathways from components $\alpha$ with influxes $I_\alpha>0$ to objective components $\Obj$.

\section{Derivation of Eqs.~(3-5) in the main text} 
\noindent\textbf{Derivation of Eq.~(\ref{eq:Slutsky_Metabolism}): Slutsky equation in microeconomics.} 
When given a utility function $u(\textbf{x})$, we define $\hat{x}_i({\bf p},I)$ as the optimal demand for good $i$ determined as a function of the price of the good ${\bf p}$ and income $I$. 
By defining $E({\bf p},u)$ as ``the minimum income required to achieve a certain utility value $u$,'' we can represent ``the minimum demand for good $i$ required to achieve a utility value $u$'' as $h_i({\bf p},u):= \hat{x}_i({\bf p},E({\bf p},u))$. 

Differentiating this function $h_i({\bf p},u)$ with respect to $p_j$ yields
\begin{eqnarray*}
\frac{\partial h_i({\bf p},\hat{u}({\bf p},I))}{\partial p_j} 
= \frac{\partial \hat{x}_i({\bf p},I)}{\partial p_j} + 
\frac{\partial \hat{x}_i({\bf p},I)}{\partial I}\frac{\partial E({\bf p},\hat{u}({\bf p},I))}{\partial p_j},
\end{eqnarray*}
where $\hat{u}({\bf p},I)$ represents the maximum utility under a given price ${\bf p}$ and income $I$. 
Due to optimality, the last term ${\partial E({\bf p},\hat{u}({\bf p},I))}/{\partial p_j}$ equals to the optimal demand $\hat{x}_j({\bf p},I)$.

Accordingly, we obtain the Slutsky equation that describes the response of demand $\hat{x}_i({\bf p},I)$ to changes in price $p_j$:
\begin{eqnarray}\label{eq:Slutsky}
\frac{\partial \hat{x}_i({\bf p},I)}{\partial p_j}=
\frac{\partial h_i({\bf p},\hat{u}({\bf p},I))}{\partial p_j}-\hat{x}_j({\bf p},I)\frac{\partial \hat{x}_i({\bf p},I)}{\partial I}.
\end{eqnarray}
The first term ${\partial h_i({\bf p},\hat{u})}/{\partial p_j}$ represents the substitution effect caused by relative changes in the price of each good; particularly, the ``self-substitution effect'' for $i=j$ is always non-positive~\cite{Varian}. 
In contrast, the second term $\hat{x}_j({\bf p},I)\frac{\partial \hat{x}_i({\bf p},I)}{\partial I}$ reflects the income effect, which can be either positive or negative. This represents the effect that an increase in the price of a good leads to an effective decrease in income, which also changes the demand for goods.

The law of mass conservation in metabolism corresponds has the perfect complementarity in economics. Accordingly, the substitution effect is always null (at a kink)~\cite{Lancaster1966,yamagishi2021microeconomics}. 
Then, by noting that $\nPath$ metabolic pathways correspond to goods in economics in our mapping between metabolism and microeconomics, we immediately obtain Eq.~(\ref{eq:Slutsky_Metabolism}) in the main text from the Slutsky equation~(\ref{eq:Slutsky}).

\noindent\textbf{Derivation of Eqs.~(\ref{eq:Quantitative_Slutsky_v}-\ref{eq:reaction_price}) from Eq.~(\ref{eq:Slutsky_Metabolism}) in the main text.} 
Noting $\hat{\vflux}=P\hat{\f}$ and multiplying Eq.~(\ref{eq:Slutsky_Metabolism}) in the main text by pathway matrix $P$ from the left, we immediately obtain equality for the responses of reaction fluxes $\vflux$: 
\begin{eqnarray} \label{eq:Slutsky_Metabolism-v_SI0}
\frac{\partial \hat{v}_i(K,{\bf I})}{\partial p^\nu_{j^\prime}}=-f_{j^\prime}(K,{\bf I})\frac{\partial \hat{v}_i(K,{\bf I})}{\partial I_\nu}
\quad(i\in\Reac,\;j^\prime\in\Path),
\end{eqnarray}

Assuming that a {stoichiometric} coefficient $S_{\mu j}$ of reaction $j$ is effectively altered to $S_{\mu j} - \Delta S_{\mu j}$, the $\jp$-th column vector of matrix $K:=SP$ changes as 
\begin{eqnarray*}
\{K_{\alpha \jp}\}_{\alpha\in\Metabo\cup\Const} \; &\to& \; \{K_{\alpha \jp} - \Delta K_{\alpha \jp}\}_{\alpha\in\Metabo\cup\Const} 
:= \{ \sum_{k\in\Reac} S_{\alpha k}P_{k \jp} - \delta_{\alpha\mu}\Delta S_{\mu j}P_{j \jp}\}_{\alpha\in\Metabo\cup\Const},
\end{eqnarray*}
with Kronecker delta $\delta_{\alpha\mu}$. 
That is, the metabolic price of pathway $\jp$ for metabolite $\mu$ changes by $\Delta p^\mu_\jp(\{\delta_{\alpha\mu}\Delta S_{\mu j}\}_{\alpha\in\Metabo\cup\Const}) = \Delta S_{\mu j}P_{j \jp}$. 

Therefore, if multiple {stoichiometric} coefficients are simultaneously altered,
\begin{eqnarray}
\Delta \hat{v}_i( \{\Delta S_{\alpha j}\}_{\alpha\in\Metabo\cup\Const})
&=&\sum_{\mu\in\Metabo}\sum_{\jp\in\Path}\frac{\partial \hat{v}_i(K,{\bf I})}{\partial p^\mu_\jp}\Delta p^\mu_\jp \nonumber \\ 
&=&\sum_{\mu\in\Metabo}\sum_{\jp\in\Path}\frac{\partial \hat{v}_i(K,{\bf I})}{\partial p^\mu_\jp}P_{j \jp}\Delta S_{\mu j}. \nonumber
\end{eqnarray}
From Eq.~(\ref{eq:Slutsky_Metabolism-v_SI0}) and $\hat{\vflux}=P\hat{\f}$, 
\begin{eqnarray}\label{eq:Slutsky_Metabolism-v_SI}
\Delta \hat{v}_i( \{\Delta S_{\alpha j}\}_{\alpha\in\Metabo\cup\Const})
&=&-\sum_{\mu\in\Metabo}\sum_{\jp\in\Path}\Delta S_{\mu j} P_{j \jp} \hat{f}_{\jp}\frac{\partial \hat{v}_i}{\partial I_\mu}. \nonumber \\
&=&-\hat{v}_j \sum_{\mu\in\Metabo} \Delta S_{\mu j} \frac{\partial \hat{v}_i}{\partial I_\mu}. 
\end{eqnarray}
Then, by defining $c^\nu_\mu(i):=\frac{\partial \hat{v}_i}{\partial I_\mu} / \frac{\partial \hat{v}_i}{\partial I_\nu}$ and
\begin{eqnarray*}\label{eq:derivation_dq}
\Delta q_i^\nu(\{\Delta S_{\mu i}\}_{\mu\in\Metabo}) := \sum_{\mu\in\Metabo}c^\nu_\mu(i) \Delta S_{\mu i}
\end{eqnarray*} 
and applying it to Eq.~(\ref{eq:Slutsky_Metabolism-v_SI}) in the case of $j=i$, we obtain 
\begin{eqnarray*}\label{eq:derivation_2}
\Delta \hat{v}_i( \{\Delta S_{\alpha j}\}_{\alpha\in\Metabo\cup\Const})
= - \hat{v}_i\frac{\partial \hat{v}_i}{\partial I_\nu}\underbrace{\sum_{\mu\in\Metabo}c^\nu_\mu(i) \Delta S_{\mu i}}_{=:\Delta q_i^\nu}.
\end{eqnarray*}
This is the derivation of Eqs.~(\ref{eq:Quantitative_Slutsky_v}-\ref{eq:reaction_price}) in the main text.

\vspace{50pt}

\vspace{30pt}

\vspace{50pt}
\begin{table*}[bth]
    \centering
    \caption{Description of the symbols in the main text} \label{table:Symbols}
    \scalebox{1}{
    \begin{tabular}{c|c}
    Symbol & Description  \\ \hline 
    $\Metabo, \Const, \Reac, \Path$ & Set of chemical species / constraints / reactions / pathways \\
    $\mathcal{E},\Obj$ & Set of exchangeable species ($\Excess\subset \Metabo$) / objective components ($\Obj\subset \Metabo\cup\Const$) \\ 
    $S,K$ & Stoichiometry matrix for reactions / pathways 
    ($S_{\alpha i},K_{\alpha \iprime}\in\mathbb{R};\alpha\in\Metabo\cup\Const, i\in\Reac,\iprime\in\Path$) \\
    $v_i,f_\iprime$ & Non-negative flux of reaction $i\,(\in\Reac)$ / pathway $i^\prime\,(\in\Path)$ \\ 
    $q^\nu_i,p^\nu_\iprime$ & Metabolic price of reaction $i$ / pathway $i^\prime$ in terms of $\nu\,(\in\Metabo\cup\Const$) \\
    $I_\alpha$ & Maximal intake of species $\alpha\,(\in\Metabo$) or total capacity for constraint $\alpha\,(\in\Const$) \\
    $\Lambda$ & Flux of objective reaction $o\,(\in\Reac)$ 
    \end{tabular}
    }
    \label{tab:symbols}
\end{table*}

\end{document}